\documentclass[prl,twocolumn,showpacs,amsmath,amssymb]{revtex4}
\usepackage{graphicx}
\usepackage{amsmath}
\input epsf

\begin{document}

\title{Time evolution in the Morse potential using supersymmetry: \\
dissociation of the NO molecule}
\author{Bal\'{a}zs Moln\'{a}r}
\author{P\'{e}ter F\"{o}ldi}
\author{Mih\'{a}ly G. Benedict}
\email{benedict@physx.u-szeged.hu}
\author{Ferenc Bartha}
\affiliation{%
Department of Theoretical Physics, Unversity of Szeged, 
Tisza L. k\"{o}r\'{u}t 84, H-6720 Szeged, Hungary}

\date{\today}
\begin{abstract}
We present an algebraic method for treating molecular vibrations in the
Morse potential perturbed by an external laser field. By the help of a
complete and normalizable basis we transform the Schr\"{o}dinger equation
into a system of coupled ordinary differential equations. We apply our method to
calculate the dissociation probability of the NO molecule 
excited by chirped laser pulses. The dependence of the molecular dipole-moment 
on the interatomic separation is determined by a quantum-chemical method, 
and the corresponding transition dipole moments are given by approximate 
analytic expressions. 
These turn out to be very small between neighboring stationary states 
around the vibrational quantum number $m=42$, therefore we propose
to use additional pulses in order to skip this trapping state, 
and to obtain a reasonable dissociation probability.
\end{abstract}

\maketitle

\affiliation{
Department of Theoretical Physics, University of Szeged, \\
Tisza Lajos k\"{o}r\'{u}t 84, H-6720 Szeged, Hungary\\}

An important model for anharmonic molecular vibrations, allowing also for
dissociation is the Morse Hamiltonian \cite{M} 
\begin{equation}
\widehat{H}_{0}=\frac{\widehat{P}^{2}}{2m}+D_{0}\left[ \exp (-2\alpha 
\widehat{X})-2\exp (-\alpha \widehat{X})\right] .
\end{equation}
We shall consider the time development of a molecular state in this
potential coupled also to an external classical field via the dipole
interaction. Thus we shall give the solution of the time dependent Schr\"{o}%
dinger equation:
\begin{equation}
i\hbar \frac{d|\Phi \rangle }{dt}=[\widehat{H}_{0}-\widehat{\mu }E(t)]
|\Phi\rangle,   \label{sch}
\end{equation}
where $\widehat{\mu} (\widehat{X})$ is the molecular dipole moment, and
$E$ is the electric field strength of the external field.
Introducing the parameters:
$\omega_{0}=\alpha \sqrt{\frac{2D_{0}}{m}}$ and 
$s=\frac{\sqrt{2mD_{0}}}{\hbar\alpha }-1/2$,
and the dimensionless operators $X=\alpha \widehat{X}$,
$P=\widehat{P}/\alpha \hbar $ the dimensionless Hamiltonian 
$H_{0}=\{(2s+1)/\hbar \omega _{0}\}\widehat{H}_{0}$
can be recast in the form 
\begin{equation}
H_{0}=A^{\dagger }(s)A(s)-s^{2}.  \label{h0}
\end{equation}
Here $A(q)=qI-(s+1/2)\exp (-X)+iP$, and $A^{\dagger }(q)=qI-(s+1/2)\exp
(-X)-iP$ are generalized supersymmetric ladder operators \cite{DKS88},
allowing the case $q\neq s$, as well. They satisfy the following commutation
relations 
\begin{eqnarray}
\left[ A(q),A^{\dagger }(q^{\prime })\right]  &=&(q+q^{\prime })I-\left(
A(q)+A^{\dagger }(q^{\prime })\right) ,\hspace{0.8cm}  \label{com} \\
\left[ A(q),A(q^{\prime })\right]  &=&0,\text{ \ \ \ \ \ \ \ \ }\left[
A^{\dagger }(q),A^{\dagger }(q^{\prime })\right] =0.  \notag
\end{eqnarray}
As $A^{\dagger }(s)A(s)=A(s)A^{\dagger }(s)-2s+1$ is a shape invariant
supersymmetric operator \cite{DKS88}, the solution of the eigenvalue
equation $H_{0}\left| \psi _{m}(s)\right\rangle =E_{m}(s)\left| \psi
_{m}(s)\right\rangle $ can be found as follows \cite{BM99} 
\begin{eqnarray}
\left| \psi _{m}(s)\right\rangle&=&\mathcal{N}_m%
A^{\dagger }(s)\cdots A^{\dagger }(s-m+1)\left| \psi _{0}(s-m)\right\rangle, 
\nonumber \\ 
E_{m}(s)&=&-(s-m)^{2}, \qquad m=0,1,\ldots \lbrack s] ,
\label{bound}
\end{eqnarray}
where $\mathcal{N}_m=\sqrt{\frac{(2s-2m)!}{m!(2s-m)!}}$, and 
$\left| \psi _{0}(s-m)\right\rangle $ is the single nondegenerate
vector annihilated by $A(s-m)$ 
\begin{equation}
A(s-m)\left| \psi _{0}(s-m)\right\rangle =0 .  \label{ralo}
\end{equation}
Using coordinate representation it can be simply proven that
$\left| \psi_{0}(s-m)\right\rangle $, as well as 
$\left| \psi _{m}(s)\right\rangle $ are
normalizable if and only if $m<s$, therefore the number of bound states is 
$[s]+1$, where $[s]$ denotes the largest integer that is smaller than $s$.
The infinite dimensional Hilbert space of the problem is, however a direct sum: 
$\mathcal{H=H}^{-}\oplus\mathcal{H}^{+}$, where $\mathcal{H}^{-}$ is the
finite ($\left[ s\right] +1$) dimensional subspace spanned by the bound
states, while $\mathcal{H}^{+}$ is an infinite dimensional subspace
with elements obtainable as continuous superpositions of
positive energy eigenstates. 

Instead of using the non-normalizable continuum energy eigenstates, 
we introduce algebraically a true orthonormal
basis in  $\mathcal{H}$  that allows a natural discretization 
of time dependent problems.  
Let $\sigma =s-[s]$, 
and starting from the state 
$\left| \phi_{0}\right\rangle $ defined by the relation
$A(\sigma )\left| \phi_{0}\right\rangle =0$,
we introduce the following series of states 
\begin{equation}
\left| \phi _{n}\right\rangle =\{{\prod_{k=1}^{n}}C_{k}^{-1}
A^{\dagger}(\sigma +k-1) \}\left|\phi_{0}\right\rangle ,  \label{pseu}
\end{equation}
where the coefficients $C_{k}=\sqrt{k(k+2\sigma -1)}$ \ ensure the
normalization. We will call the states $\left| \phi _{n}\right\rangle $
quasi-number states \cite{pss}. From the definition (\ref{pseu}) it is also
seen that the neighboring quasi-number states are connected as 
\begin{eqnarray}
A^{\dagger }(\sigma +n)\left| \phi _{n}\right\rangle &=& C_{n+1}\left| \phi
_{n+1}\right\rangle,  \nonumber \\
A(\sigma +n)\left| \phi _{n}\right\rangle
&=& C_{n}\left|\phi _{n-1}\right\rangle.  \label{quasigen2}
\end{eqnarray}
The orthogonality of the $\left| \phi _{n}\right\rangle $ states can be seen
by using the commutators (\ref{com}). In coordinate representation they have
the form
\begin{equation}
\phi _{n}(x)=\sqrt{{n!}/{\Gamma(2\sigma +n)}} y^{\sigma }(x)\exp (-%
\frac{y(x)}{2})L_{n}^{2\sigma -1}(y(x)) , \label{pseuwav}
\end{equation}
where $L_{n}^{2\sigma -1}$ is a generalized Laguerre polynomial 
\cite{ASHMF,Szego} of the variable $y(x)=(2s+1)e^{-x}$. 
In contrast to the energy eigenfunctions, these wave functions 
form a complete orthonormal set of square integrable functions 
\cite{Szego}, therefore the quasi-number states 
constitute a true orthonormal basis in the
Hilbert space. 

Using the properties of the supersymmetric 
ladder operators $A^{\dagger }$ and $A$, one can
calculate the matrix of $H_{0}$ in this basis
\begin{eqnarray}
\left\langle \phi _{m}\right| H_{0}\left| \phi _{n}\right\rangle =
\left(C_{m}^{2}-s^{2}+(m-\left[ s\right] )^{2}\right)
\delta _{m,n} \nonumber \\ 
+(\left[ s\right] -n)C_{m}\delta _{m,n+1}+
(\left[ s\right] -m)C_{n}\delta_{m+1,n} .
\label{hmat}
\end{eqnarray}
It consists of two blocks, the one corresponding 
to $0\le~n,m~\le [s]$  
is $\left[ s\right]+1$ dimensional, whereas the other one,
corresponding to $n,m >[s]$ 
is infinite dimensional, and both blocks are tridiagonal.
The block structure shows that the first $\left[ s\right]+1$
quasi-number states span an invariant subspace of $H_0$,
in which it can be diagonalized.
This subspace should coincide with  $\mathcal{H}^{-}$, as $H_0$
has exactly $\left[ s\right] +1$ nondegenerate eigenvalues
with normalizable eigenstates.
The orthogonality relation 
\begin{equation}
\left\langle \phi _{n}\right| \psi _{m}\rangle =0,\text{ \ \ for }%
m=0,1\ldots \left[ s\right] ,\text{\ \ and }n>\left[ s\right] \text{\ \ }
\label{ort}
\end{equation}
equivalent to the above statement can be proven directly, and the expressions 
of the overlaps $\left\langle \phi _{n}\right| \psi _{m}\rangle$
for $n \le [s]$ can be determined explicily \cite{MBtobe}. 

We will consider the time evolution of molecular vibrations subject to an
external classical field as described by Eq. (\ref{sch}).
In the dimensionless units used in Eq. (\ref{h0}) the
interaction term,
$\widehat{H}_{\text{int}}=-\widehat{\mu }(\hat{X})E(t)$
can be written as 
\begin{equation}
H_{\text{int}}=-\mu (X)\mathcal{E}(t),  \label{hint}
\end{equation}
where we have introduced $\mu =\frac{\alpha }{q_{e}}\widehat{\mu }$, with an
effective charge $q_{e}=\frac{d\widehat{\mu }(x)}{dx}|_{x=0}$, and $\mathcal{%
E}(t)=\frac{(2s+1)q_{e}}{\hbar \omega _{0}\alpha }E(t)$ denotes the electric
field strength in our units. As we will solve the Schr\"{o}dinger equation (%
\ref{sch}) in the quasi-number state basis, we also need to know the matrix
of the dipole moment in this basis. Due to the algebraic properties of the
quasi-number states, the matrix of $\mu (X)$ can also be determined
analytically for certain cases. Let us suppose that the dipole
moment can be approximated by an operator of the form 
\begin{equation}
\mu(X)={\textstyle{\sum_{i} \mu_{i}(X)=
\sum_{i} (a_{i} X+d_{i}) e^{-\gamma_{i} X}}} ,  \label{dip}
\end{equation}
where $a_i$, $d_i$ and $\gamma_{i}$ are real numbers.
Taking only a single term with $d=\gamma =0$ 
would mean the simplest linear dipole moment borrowed from
atomic calculations and used sometimes  
for molecules  as well. 
This latter approach, however, overestimates the 
strength of the interaction at larger atomic
separations and leads to an unrealistically high 
dissociation probability.  

By the aid of Eq. (\ref{pseu}) and the fact that
$\left[ \mu(X),A^{\dagger }(q)\right] =
\left[ \mu(X),-iP\right] =\frac{d\mu (X)}{dX}$,
recurrence relations can be derived for the matrix elements of $\mu(X)$
between the states $|\phi_{n}\rangle$. 
We have 
\begin{eqnarray}
\langle \phi _{m}| Xe^{-\gamma X}|\phi _{n+1}\rangle =
[(n-m-\gamma)\left\langle \phi _{m}\right|Xe^{-\gamma X}\left| \phi_{n}
\right\rangle \nonumber \\ 
+ C_{m}\left\langle\phi _{m-1}\right|Xe^{-\gamma X}\left| \phi _{n}\right\rangle
+\langle \phi _{m}|e^{-\gamma X}| \phi _{n}\rangle]/
{C_{n+1}},  \nonumber
\end{eqnarray}
\begin{eqnarray}
\left\langle \phi _{m}\right| e^{-\gamma X}\left| \phi
_{n+1}\right\rangle =[C_{m}\left\langle \phi _{m-1}\right|
e^{-\gamma X}\left| \phi _{n}\right\rangle )  \nonumber  \\ 
+(n-m-\gamma )\left\langle \phi _{m}\right|e^{-\gamma X}
\left| \phi _{n}\right\rangle]/{C_{n+1}}.  
\label{dip2mat}
\end{eqnarray}
Then all the necessary matrix elements can be calculated, 
starting from terms 
$\langle \phi _{0}|X \exp(-\gamma X)| \phi _{0}\rangle $ and
$\left\langle \phi _{0}\right| \exp(-\gamma X)\left| \phi
_{0}\right\rangle $. These latter can be obtained via integration in
coordinate representation using the wave function $\phi _{0}(x)$ 
of Eq. (\ref{pseuwav}), and we get 
\begin{equation*}
\left\langle \phi _{0}\right|Xe^{-\gamma X}\left| \phi _{0}\right\rangle =
\frac{\Gamma (2\sigma +\gamma)}{\left( 2s+1\right)^{{\gamma}}\Gamma (2\sigma )%
}[\ln (2s+1)-\widetilde{\psi }(2\sigma +\gamma)],
\end{equation*}
\begin{equation}
\left\langle \phi _{0}\right| e^{-\gamma X})\left|\phi_{0}\right\rangle =%
\frac{\Gamma (2\sigma +\gamma)}{\left( 2s+1\right)^{{\gamma}}
\Gamma(2\sigma)},   \label{rec0}
\end{equation}
where $\widetilde{\psi }$ denotes Euler's digamma function \cite{ASHMF}.

The values of the parameters 
$a_{i}$, $d_i$ and $\gamma _{i}$
can be obtained by fitting the function (\ref{dip})
to experimental dipole curves, or to those 
obtained from molecular calculations.

The quasi-number states constitute a complete orthonormal basis,
hence one can expand the solution
$\left| \Phi (t)\right\rangle $
of Eq. (\ref{sch}) in the form:
$|\Phi (t)\rangle ={\sum_{n=0}^{\infty }}c_{n}(t)| \phi _{n}\rangle$.
Measuring the time $t$ in units of $T=2\pi /\omega _{0}$,
the Schr\"{o}dinger equation can be written as the
following infinite system of ordinary differential equations
\begin{equation}
i\dot{c}_{n}=\frac{2\pi}{2s+1}
{\sum_{m=0}^{\infty }}\left\{ \langle \phi
_{n}| H_{0}| \phi _{m}\rangle -\mathcal{E}(t)\langle
\phi _{n}| \mu | \phi _{m}\rangle \right\} c_{m}.
\label{dyn}
\end{equation}
where $\langle \phi _{n}| H_{0}| \phi _{m}\rangle $
and $\langle \phi _{n}| \mu | \phi _{m}\rangle $ are
given by the expressions (\ref{hmat}) and 
(\ref{dip2mat}),
respectively.
We note here that the usual method of solving the time evolution 
of the molecular state is based on the split operator method \cite{FFS82}.
That approach solves the partial differential equation corresponding
to Eq. (\ref{sch}), 
and yields an approximate wave function of the variables $x$ and $t$.
In contrast, we solve a system of ordinary equations allowing a higher accuracy
in the calculations.  

We will apply our method to determine the dissociation probability of a
molecule. This is a time dependent quantity defined as the projection
probability of $\left| \Phi (t)\right\rangle $ on $\mathcal{H}^{+}$ 
that can be given as
\begin{equation}
\mathcal{P}(t)=1-{\sum_{m=0}^{[s]}} | \langle \Phi(t)| \psi _{m}\rangle |^{2}
=1-{\sum_{m=0}^{[s] }}c_{m}^{\ast }(t)c_{m}(t). 
\label{disprob}
\end{equation}
To calculate the dissociation probability above, one has to solve the set of
the dynamical equations (\ref{dyn}) and determine the coefficients 
$c_{m}(t)$ for $0\leq m \leq [s]$. 
In practice one has to truncate the system (\ref{dyn}), and we have to use a
numerical method to find the solution. 

The restriction of the number of the
dynamical equations to a finite number is done here in two steps. First we
consider the problem in a large subspace of dimension $N\gg [s],$
and neglect the contribution of those states $|\phi_{n}\rangle $
for which $n>N$ .
Then the Hamiltonian is represented by an operator which is restricted 
to the finite $N$ dimensional
subspace, spanned by the orthonormal system: 
$\{\left| \psi _{m}\right\rangle ,\text{\ }m=0,1,2\ldots \left[ s\right] ,%
\text{\ \ }\left| \phi _{n}\right\rangle ,\text{\ }n=\left[ s\right]
+1,\ldots \ N\}. $
We denote this operator by $H^{(N)}=H_{0}^{(N)}-\mu ^{(N)}E(t)$ where 
$H_{0}^{(N)}$ and $\mu^{(N)}E(t)$ are the truncated operators, so that their
matrices are of dimension $N$. 

Before solving the already finite number of ordinary
differential equations, we perform a unitary transformation on our truncated
basis and bring $H_{0}^{(N)}$ into diagonal form for $n>[s]$, too.
Due to the fact that the matrix of $H_{0}^{(N)}$ is finite tridiagonal
this can be done by the help of a very fast algorithm.
This yields positive energy eigenvalues
$E_{n}^{(N)}$ ($\left[s\right] < n\leq N$) and 
eigenstates $\left| \psi _{n}^{N}\right\rangle $, which we can
use for the description of the dissociated  molecule. If $N$ is
large enough, then the lowermost positive energies $E_{n}^{(N)}$ follow
densely each other, and approximate satisfactorily the continuous energy
spectrum above the dissociation threshold. In the second step of the
approximation, we restrict the calculation to the bound subspace and to 
those  $\left| \psi _{n}^{N}\right\rangle$  states with $n>[s]$, for which the 
dipole couplings with the bound energy eigenstates are non-negligible.  
In the application to be discussed below, it 
turns out that the latter are significant only for those  
states that correspond to the lowest positive energy values $E_{n}^{(N)}$ 
with $n=[s] +1,\ldots \ M$, so that $M \ll N$.
Therefore, in order to follow the
dissociation process, it is enough to solve Eq. (\ref{dyn}) in the basis 
$| \psi _{m}\rangle $, $| \psi _{n}^{N}\rangle$, which
means that we expand the time dependent state as 
$| \Phi(t)\rangle =\sum_{n=0}^{M}b_{n}(t)| \psi _{n}\rangle $,
where now the notation $| \psi _{n}\rangle $ has been used also for
$| \psi _{n}^{N}\rangle $ with 
$n=[s]+1, \ldots M$. 
Then we solve the system of $M$ ordinary differential equations
that describes the time evolution of the coefficients $b_{n}$
and calculate the dissociation probability via Eq. (\ref{disprob})
with the obvious $c_{m}(t) \rightarrow b_{m}(t)$ replacement. 

In what follows we are going to apply our method for
the determination of the dissociation probability 
of the nitrogen-oxide (NO) molecule, excited 
by an appropriate laser field \cite{T94}. 
We have used a Morse potential with the following data 
in the electronic ground state \cite{HH}:
$m=7.46\ {\mathrm{a.u.}}$, 
$D=6.497\ {\mathrm {eV}}$ and  $\alpha=27.68\ {\mathrm{nm^{-1}}}$,
yielding $s=54.54$. 
\begin{figure}[htbp]
\includegraphics*[bb=80 40 820 505 ,width=8.25cm]{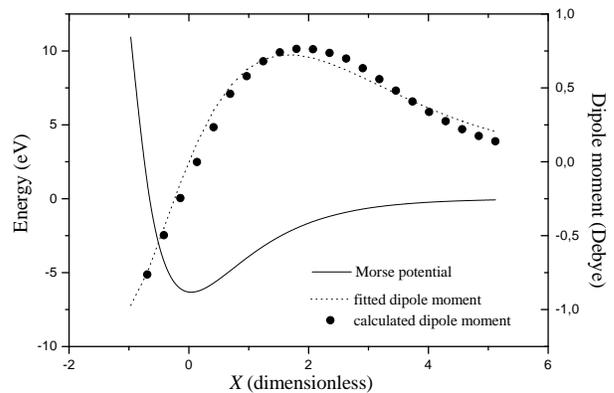}
\caption
{
The Morse potential and the dipole moment of the NO molecule.
}
\label{potdip}
\end{figure}
The potential and the dependence of the dipole moment on the interatomic
separation are depicted in Fig. \ref{potdip}.
The circles show the dipole moments of NO at various
interatomic separations that we have calculated by an
unrestricted density functional method \cite{BF}.
The dotted line is a fitted function of the
form Eq. (\ref{dip}) with  two terms, 
and with the parameter values: $a_{1}=-9.66$, $a_{2}=10.64$, 
$d_1=d_2=0$, $\gamma _{1}=0.927$, $\gamma _{2}=0.870$. 

\begin{figure}[htbp]
\includegraphics*[bb=45 50 760 520 ,width=8.25cm]{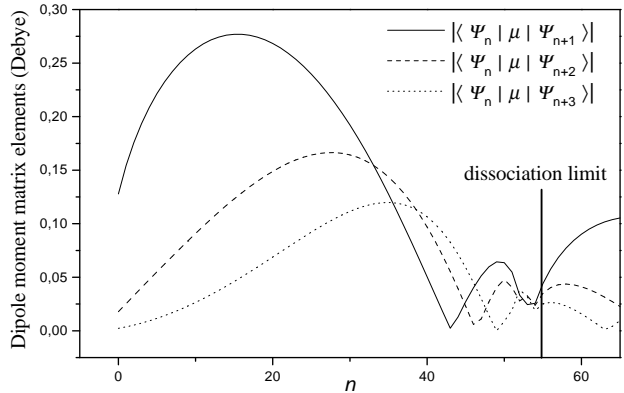}
\caption
{
Absolute values of the dipole matrix elements between first,
second, and third neighbor states. 
The matrix elements far from the diagonal are diminishingly small.  
\label{dips}
}
\end{figure}

The corresponding matrix elements 
have been determined analytically by using the
recurrence relations (\ref{dip2mat}-\ref{rec0}),
and those between the first second and third neighbors
are shown in Fig \ref{dips}. 
In our calculations we have chosen the truncation
indices to be $N=3000$, and $M=200$, as increasing these values
had no effect on the results. 
(Note that the number of bound states is now $[s]+1=55$). 
The initial state was taken to be
the vibrational ground state, and we devised the excitation with the 
aim to acheive a significant dissociation probability. 

\begin{figure}[htbp]
\includegraphics*[bb=75 55 795 505 ,width=8.25cm]{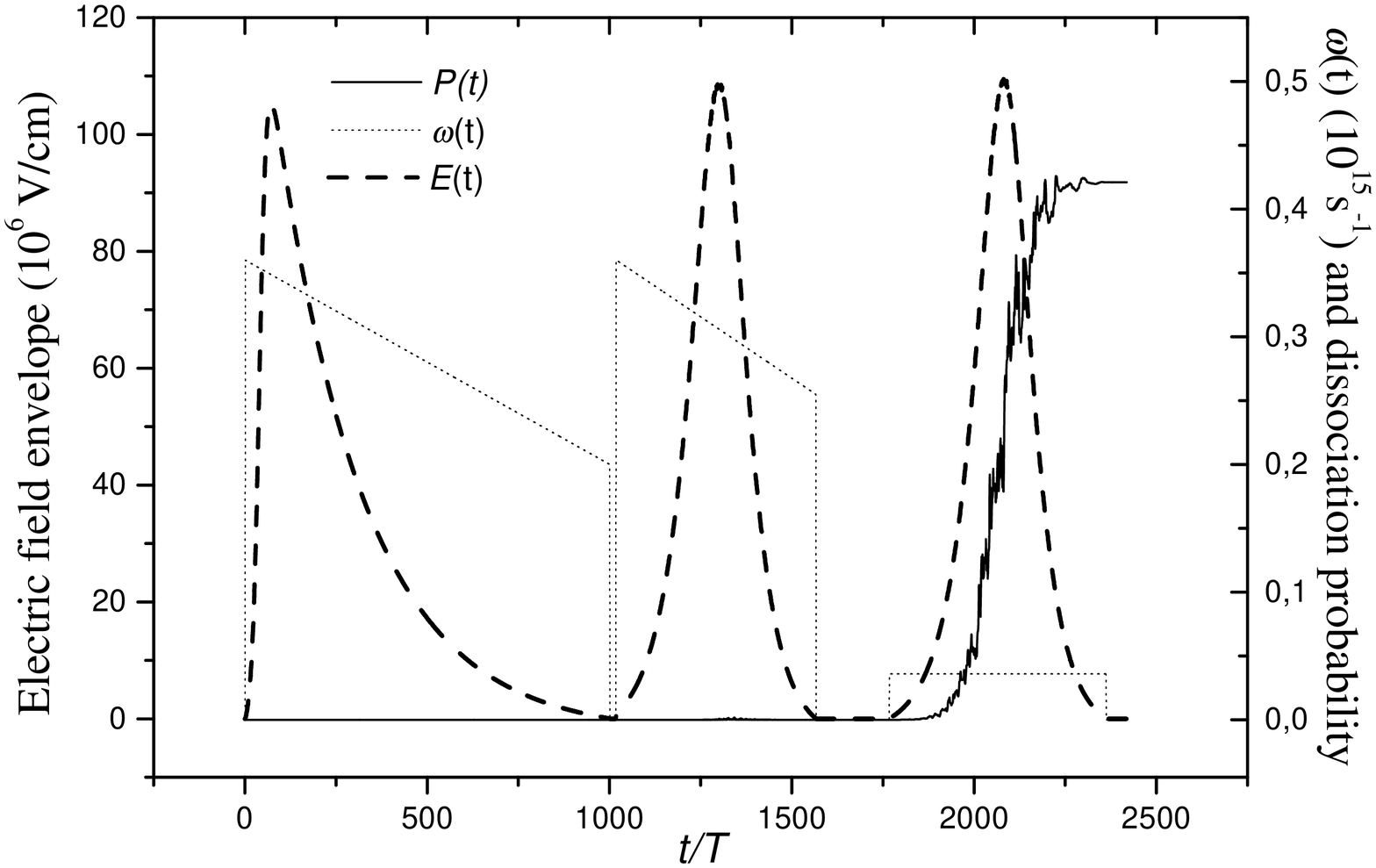}
\caption
{
Electric field strength envelopes and time dependent frequencies of the 
applied pulses. Dissociation probability is also shown as a function of
time measured in units of $T=2\pi/\omega_{0}$. All three pulses 
have a distorted secant-hyperbolic envelope (see text) with the data: 
pulse (1): $\tau_f=10.27$, $\tau_t=246.4$, $t_0=41.0$; 
pulse (2): $\tau_f=71.87$, $\tau_t=61.61$, $t_0=1297.5$;
pulse (3): $\tau_f=71.87$, $\tau_t=61.61$, $t_0=2087.6$.
Peak values for all three pulses are: $1.05\times 10^{8}\ \mathrm{V/cm}$.
\label{pulses}
}
\end{figure}

According to \cite{CBC90} one can think of the desired
time evolution as a sequence of two-level transitions.
There are two conditions to be fulfilled: the  
electric field should be approximately resonant with the 
actual transition, and its area per transition \cite{CBC90} 
should be around $\pi$. 
In order to maintain the resonance condition during the whole process, 
it is appropriate to use continuously chirped pulses \cite{CBC90}.
If the frequency of the pulse decreases slowly enough, 
then it is possible to consider its effect 
as a sequence of approximately resonant two-level transitions.
For low vibrational quantum numbers the coupling is largest 
between the nearest neighbors, therefore at the beginning  the 
excitation has to drive the system through the
ladder of neighboring energy eigenstates
towards the continuum. However, as it is seen in Fig. \ref{dips}, 
there is  a minimum in the $m \rightarrow m+1$ 
couplings at $m=42$, 
therefore this vibrational quantum number corresponds to a trapping state. 
This is a consequence of the shape of the dipole moment curve
(Fig. \ref{potdip}) and it would be absent in the
linear approximation $\mu(X) \sim X$.
One can circumvent the problem by lengthening the pulse in time,
but we present a different route here with the three pulses
shown in Fig. \ref{pulses} exploiting other
than first neighbor transitions, as well.  
The area condition could be met for all three pulses 
with distorted secant-hyperbolic envelopes:
$\mathcal{E}=\mathcal{E}_{a}/[\exp(t_0-t)/\tau_{f}+\exp(t-t_0)/\tau_{t}]$
with different exponentials in their fronts $(\tau_{f})$, 
and tails $(\tau_{t})$.  
The chirping rates are seen in Fig. \ref{dips}, while 
the other data of the pulses are given in the caption. 
Using these parameters, 
the first pulse drives the molecule into a superposition of
a few states $| \psi _{m}\rangle $ around $m=31$. 
As seen in Fig. \ref{dips}, the $m\rightarrow m+2$ matrix elements 
are getting larger here than those connecting the neighboring states, 
and  the former have also nonvanishing values around $m=42$. 
This implies that a second laser pulse with an initial carrier frequency
being in approximate resonance with the corresponding $m\rightarrow m+2$
transition will continue the dissociation process and let the 
molecular state jump over the trap. 
After this second pulse, the distribution of the populated states is not so
narrow as it was before, which is a consequence of the presence  
of second neighbor transitions. 
The most probably populated states are now  
so close to the continuum limit that a final pulse 
-- even without chirping --
leads to a dissociation probability that is more than $40 \%$. As shown
in Fig. \ref{pulses}, this probability is practically
zero before the third pulse. 

In summary, we have presented a method based on supersymmetric 
quantum mechanics that leads to a system of ordinary 
differential equations describing molecular time evolution. We have applied 
this method for devising laser pulses leading to a significant
dissociation probability of the NO molecule with a realistic dipole 
moment.  

We thank A. Czirj\'{a}k and A. L\H{o}rincz for discussions. This work was
supported by the Hungarian Scientific Research Fund (OTKA) under contract
No. T32920, and by the Hungarian Ministry of Education under contract 
No. FKFP 099/2001.


\begin{thebibliography}{99}
\bibitem{pss} A somewhat different set of basis functions
called pseudo-number states were introduced in Ref. \onlinecite{BM99} 
in the construction of the coherent states of the Morse potential.
Time evolution in terms of pseudo-number states was considered in 
Ref. \onlinecite{MBF01}, while the algebraic background 
has been discussed in Ref.  
\onlinecite{MBB01}.
\bibitem{MBtobe} These calculations use the properties of
the supersymmetric ladder operators, and will be published elsewhere.
\bibitem{BF}  We have used 
the Gaussian 98 program package with the B3LYP
functional and the 6-311G(3d2f) basis set.
\end{thebibliography}
\end{document}